\documentstyle[psfig]{l-aa}
\def\eg{e.\ g.} 
\def\etal{et al.} 
\def\apgt{\ {\raise-.5ex\hbox{$\buildrel>\over\sim$}}\ }
\def\aplt{\ {\raise-.5ex\hbox{$\buildrel<\over\sim$}}\ }

\newcommand{\pyr}{\mbox {{\rm yr$^{-1}$}}}

\newcommand{\msun}{\mbox{${\rm M}_\odot$}}

\newcommand{\nsns}{\mbox {${\rm (ns,~ns)}$}}
\begin{document}

\thesaurus{08.02.1; 08.05.3; 08.14.1; 12.03.4; 13.07.1; 13.07.3}
\title{Gamma-ray bursts and density evolution of neutron star binary mergers}
\author{Philippe Bagot\inst{1}, 
        Simon F. Portegies Zwart\inst{2} and 
        Lev R. Yungelson\inst{3,4}}
\institute{Graal, Upresa 5024/CNRS, Universit\'e Montpellier II,
	   CC072, F-34095 Montpellier,
	   France; bagot@graal.univ-montp2.fr  
\and
	   Astronomical Institute {\em Anton Pannekoek}, 
           Kruislaan 403, 1098 SJ Amsterdam, 
	   The Netherlands; spz@astro.uva.nl
\and
	   Institute of Astronomy of the Russian Academy of Sciences, 
	   48 Pyatnitskaya Str., 109017 Moscow, 
           Russia; lry@inasan.rssi.ru
\and
           DARC, Observatoire de Paris, Section de Meudon,
           92190 Meudon Cedex, France}

\date{Received date/ Accepted date}
\maketitle
\markboth{P.~Bagot \etal:\,\, $\gamma$-bursts}{}

\begin{abstract}
The evolution of the comoving cosmic merger-rate density of neutron
star binaries $n_c(z)$ is calculated using a distribution of their
merging times provided by population-synthesis computations of binary
stars.  We adopt an exponential law for the star formation rate with
various timescales for different morphological types of galaxies. For
elliptical galaxies also an initial burst of star formation, lasting
one Gyr, is considered.  The resulting $n_c(z)$ of most models
agree with the form $n_c(z) \propto (1+z)^{1.5-2}$ for $ z \aplt 2$,
which has been proposed for the source population of $\gamma$-ray
bursts.  Assuming a standard candle luminosity, the computed
brightness distribution is consistent with the BATSE results if bursts
at the peak flux threshold, $P$~=~0.4 photons~cm$^{-2}$~s$^{-1}$, are
located at a limiting redshift of 1.9 to 3.3. 
Progenitors of the systems producing $\gamma$-ray bursts at small
redshift (bright) are likely to host in spiral galaxies and star
forming regions whereas these at high redshift (dim) reside mainly in
elliptical galaxies.
The location of a burst may be up to $\sim 1$\,Mpc away
from the host galaxy.

\end{abstract}

\section{Introduction}

The recent discoveries of optical counterparts of the $\gamma$-ray
bursts GRB~970228 (Groot \etal\ 1997)
and
GRB~970508 (Bond 1997) and the measured emission-line redshift of
$z=0.853$ (Metzger et al.\ 1997) for the latter provide evidence for
their cosmological origin.  A possible source for $\gamma$-ray bursts
(GRBs) may be the merging of two neutron stars in a close binary
(Blinnikov \etal\ 1984).

For cosmologies with no vacuum energy, the brightness distribution of
burst intensities expected for a uniform source population is
consistent with the BATSE distribution if the limiting redshift is
about unity (e.g. Dermer 1992; Mao \& Paczy\'nski 1992).  However,
cosmological time-dilation effects in the BATSE sample indicate that
the dimmest sources should be located at $z \approx 2$ (Norris
et al. 1995).  If this is the actual limiting redshift,
a source population with a
comoving rate density of the form $n_c(z) \propto
(1+z)^{\beta}$ is compatible with the BATSE distribution for $1.5
\aplt \beta \aplt 2$ (Horack et al. 1995).
Therefore, it is worthwhile to examine whether
the comoving merger rate density of neutron-star binaries evolves in a
similar fashion.  

By means of population synthesis computations for binary stars the
merger rate of neutron-star binaries [hereafter \nsns] can be computed
(see Portegies Zwart \& Yungelson 1998 and references therein).  Using
this approach Lipunov \etal\ (1995) computed the evolution of \nsns\
mergers as a function of redshift and $\log N - \log S$\ distributions
of GRBs for a cosmic population which contains galaxies with a
constant star-formation rate and also galaxies with initial bursts of
star formation in different proportions.  Totani (1997) computed the
evolution of the GRBs rate density from \nsns\ mergers in a model
based on the observationally determined history of cosmic star
formation and in a model derived from detailed galaxy evolution.
Totani assumed a distribution of merging times $f(t_c)\propto
t_c^{-1}$ for \nsns\ systems\footnote{Sahu \etal\ (1997) made similar
computations for ``cosmic'' star formation history and a fixed merger
time of $3 \times 10^7$\ yr.}.
We go a step further by adopting for
all galaxies exponentially decreasing star formation rates (SFR) with
different timescales depending on galaxy morphology (\eg\ Sandage
1986). We apply the distribution of $t_c$ from model B of
Portegies Zwart \& Yungelson (1998, henceforth PZY98) which provides
the best fit to the expected birthrate and orbital parameters of the
Galactic population of high-mass binary pulsars.

\section{Model}
The annual \nsns\ birthrate is proportional to the SFR in
the galaxy $\Psi(t, \tau),$\ where $\tau$ is the timescale for star
formation. 
The SFR adopted is a decreasing exponential, 
which appears when the SFR is supposed to be proportional to
the gas content without taking into account the gas ejected by stars
(e.g.\ Bruzual \& Charlot 1993): 
\begin{equation}
\Psi(t_{\rm gal},\tau)=\tau^{-1} \exp (-t_{\rm gal}/\tau),
\label{expdec}
\end{equation}
where the age of the galaxy is given by $t_{\rm gal}=t-t_F$ and $t_F$\
the time at which the galaxy was formed.  At any time $t$ the \nsns\ merger
rate $\mu(t, \tau)$ has a contribution from systems
that are formed at different epochs $t-t_c$ in the history of that
galaxy.  It can thus be expressed as a convolution integral of the
birthrate of stars $\Psi(t, \tau)$\ and the distribution of \nsns\
merging times $f(t_c)$:
\begin{equation}
\label{nuc}
\mu(t,\tau)=\mu_\circ \int^{t-t_F}_0 \!\! f(t_c)~\Psi(t-t_c,\tau)~dt_c,
\end{equation}

The results of the population synthesis computations give $f(t_c)$\
and the normalization coefficient $\mu_0$.  The function $f(t_c)$\
from PZY98 is approximated by a Gaussian with $x=\log (t_c/t_{\rm
gal}(0))$ as a parameter, the maximum at $x_\circ=-2$ and with
$\sigma$=1 ($t_{\rm gal}(0)$\ is the age of the galaxy at $z = 0$).
We assume that the distribution function $f(t_c)$, normalized to
unity, is time-independent and the same for all galaxies.  For the
normalization of $\mu(t, \tau)$\ we require that in a reference galaxy
similar to the Milky Way (type Sb, $ M_{\rm ref} = 2 \times 10^{11}\,
\msun$, with a current astration rate of $4\, \msun\,{\rm yr^{-1}}$)\
the current rate of \nsns\ mergers is $2 \times 10^{-5}\, \pyr$, which
is the merger rate obtained for the Galaxy by PZY98.

For simplicity, we split the Hubble sequence into three types of
galaxies: E to SO, Sa to Sb and Sc to Sd.  For our selected mixture of
Hubble types, two sets of characteristic star formation timescales are
used (cf.\ Table 1). This parameterization is rather simplistic; the
fraction of E/S0 galaxies may differ considerably from 20\% (Dressler
1980) and the star formation history in spirals of the same
morphological type may be a function of their mass (Gallagher \etal\
1984).  If indeed most stars in the Universe formed in dwarf
star-burst galaxies (\eg\ Babul \& Ferguson 1996) at $z \sim 1$, this
may affect our results considerably.

Star formation is assumed to occur continuously according to
Eq.~\ref{expdec} in all galaxies. We also investigate the case of an
initial burst of star formation during the first Gyr in E-SO galaxies
and no star formation thereafter. The latter models are denoted as
sets~1b and~2b.

\begin{table}[t]
\caption{ Adopted parameters for galaxies of different morphological
types: timescale of star formation $\tau_i$; contribution to the
$B$-band luminosity of the Universe $c_i$ (Phinney 1991); mass to
blue-light ratio $M/L_B$ (after Lipunov \etal\ 1995 and Guiderdoni \&
Rocca-Volmerange 1987).  }
\begin{tabular}{lrrcc}
\noalign{\smallskip}
\hline
\noalign{\smallskip}
Type & \multicolumn{2}{c}{$\tau_i$ (Gyr)} & 
       $c_i$ & $(M/L_B)_i$ \\
     & set~1 & set~2 &  & [$h\times(M/L_B)_{\odot}$]\\
\noalign{\smallskip} \hline
\noalign{\smallskip}
E - SO  & 1  & 1   & 20 \% & 10 \\
Sa - Sb & 4  & 6   & 40 \% & 5  \\
Sc - Sd & 10 & 15  & 40 \% & 2  \\
\noalign{\smallskip}
\hline
\end{tabular}
\end{table}

The comoving rate density can be related to $\mu(z,\tau)$ via the
B-band luminosity density from the Universe $L_B=1.9 \times
10^8~h~L_{\odot B}~{\rm Mpc^{-3}}$ (Efstathiou et al.\ 1988) where
$h=H_\circ/100\,{\rm km\, s^{-1} Mpc^{-1}}$:
\begin{equation}
n_c(z)=\frac{L_B}{M_{\rm ref}} ~\sum_i ~c_i \left(\frac{M}{L_B}\right)_i
        \mu(z,\tau_i).
\end{equation}
Here the summation is taken over all morphological types and $c_i$ is
the contribution of each type of galaxy to the B-band luminosity of
the Universe, $M/L_B$ is in solar units (see Table~1).

The comoving rate density depends on the cosmological parameters
$(h,~\Omega_\circ,~\lambda_\circ)$ and on the redshift of formation
$z_F$, via the relation for the age of 
galaxies relative to redshift: $t_{\rm gal}(z)=t(z)-t(z_F)$. 
We consider two cosmological models with $z_F=5$
and no vacuum energy: $(h,~\Omega_\circ)=(0.5,~1.0)$ and (0.75,~0.2)
for which $t_{\rm gal}(0)$=12.16 and 9.92~Gyr, respectively. 

\section{Application}
Figure~1 shows the redshift dependence of $n_c(z)$ for several of our
models. We normalize the merger rate density to unity at $z$=0.  The
gray shaded area demonstrates that the results are basically
consistent with the estimates derived from the GRB source distribution
as observed by BATSE (Horack et al.\ 1995); the computed merger rate
falls roughly in the allowed region for $z\la 2$.  Qualitatively, the
behavior of $n_c(z)$ is easily understood from Fig.~2 which shows the
relative contribution of galaxies of different types to the \nsns\
merger rate (for set 1).  At high $z$\ it is dominated by early-type
galaxies with initially a high time-averaged SFR.  In set Ib (upper
solid line) the sudden drop at $z \sim 2.5$ reflects the cessation of
star formation in ellipticals in one Gyr after formation and the
relatively short typical \nsns\ merger time of $\aplt 1$\ Gyr (PZY98).
The more gradual decline of $n_c(z)$\ in set 1 reflects the continuous
star formation in both elliptical and spiral galaxies. Comparison of
sets 1 with 2 suggests that the evolution of $n_c(z)$\ is better
described by the models with longer star-formation time scales.  The
lower solid line uses the same merger-time function $f(t_c)$ as Totani
(1997). This model predicts a lower \nsns\ merger rate.
 
\begin{figure}[t]
\centerline{
\psfig{file=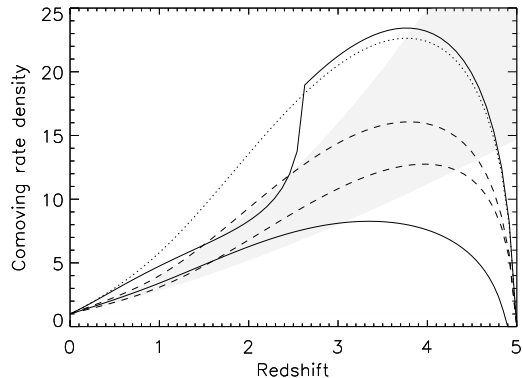,bbllx=95pt,bblly=375pt,bburx=530pt,bbury=695pt,height=5cm}}
\caption[t]{ Evolution of the comoving \nsns\ merger rate density as a
function of redshift, normalized to unity at $z$=0. The shaded area
corresponds to the allowed region according to $n_c(z) \propto
(1+z)^{\beta}$ with $\beta$ in the range 1.5~to~2. The upper solid
(dotted) line is for set 1b (set 1) with $t_{\rm
gal}$(0)=12.16~Gyr. The dashed lines give the rates for set 2 with
$t_{\rm gal}$(0)=12.16~Gyr (upper curve) and 9.92 Gyr (lower curve).
The lower solid line has to be compared with the dotted line and
corresponds to the merger time distribution $f(t_c) \propto t_c^{-1}$
with a lower cut-off at 0.02 Gyr (like in Totani 1997).}
\end{figure}

\begin{figure}[t]
\centerline{
\psfig{file=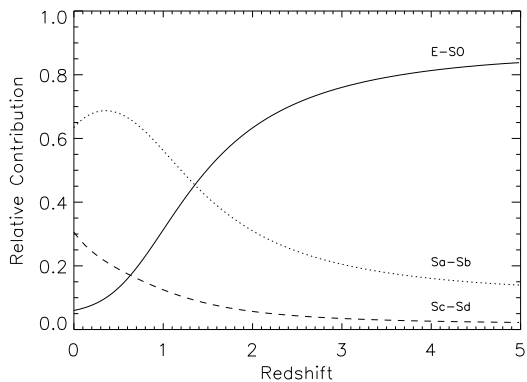,bbllx=95pt,bblly=375pt,bburx=530pt,bbury=695pt,height=5cm}}
\caption[t]{Relative contribution of galaxies of different types to
the \nsns\ merger rate density. The current galactic age is $t_{\rm
gal}(0)$=12.16 Gyr and set 1 of the SF timescales is used.}
\end{figure}

The upper and lower dashed lines in Fig.~1 (set~2) demonstrate the
effect of the age of galaxies on the comoving merger rate, i.\ e.; the
influence of the cosmological parameters. 

Figure~2 gives the relative contribution to the \nsns\ merger-rate
density for each selected subclass of galaxies. The majority of events
which are potentially detectable by gravitational-wave observatories
(LIGO/VIRGO) are located in early-type spiral galaxies. If GRBs
originate from \nsns\ coalescence the dimmest bursts are expected to
be hosted in elliptical galaxies.

Following the standard procedure (e.g. Horack et al. 1996) we compute
the number of bursts $N(>P)$ with a peak flux greater than $P$.  We
assume that bursts are standard candles and the intrinsic luminosity
doesn't evolve.  The spectral form of the burst similar to that
observed is adopted: $\Phi(E) \propto E^{-1} \exp(-E/E_0)$, where
$E_0$=350 keV is a characteristic energy.  For comparison, we use the
observed integral brightness distribution from the BATSE 3B catalog in
the energy range 50-300 keV measured at a timespan of 1024 ms (Meegan
et al.~1996). Figure 3 shows the expected brightness distributions
computed for sets of parameters ~1 and ~1b, superimposed on the BATSE
data.  The curves are normalized at the peak flux threshold $P$=0.4
photons~cm$^{-2}$~s$^{-1}$. Figure~4 provides the results of a
Kolmogorov-Smirnov test of the BATSE 3B catalog to the results of our
computations.  Only data above the peak flux of 0.4
photons~cm$^{-2}$~s$^{-1}$ are used to avoid threshold effects.  The
highest confidence level (CL)~is obtained for sets~1 and ~1b if the
limiting redshift $z_{0.4}$ at the peak flux threshold is $\sim$ 2.4
and 3.0 respectively.  For other values of the star formation
timescales and cosmological parameters $z_{0.4}$ is found to range
from 1.9 to 2.7 (with a CL$~>80\%$) for the models with a burst of
star formation in elliptical galaxies (models 1b and 2b).  Without an
initial burst of star formation $z_{0.4}$ ranges from 2.9 to 3.3.
Similar results are found for $E_0$ in the range 300~-~400 keV.

Note that for higher limiting redshift, Fig.4 shows the existence of
other possible fits with a lower CL (i.e. model 1b,
$z_{0.4}=3.8$). The primary peak at $z_{0.4}=2.4$ corresponds to the
first change of the slope of the comoving rate density 
(Fig.1). For higher redshifts, the sudden increase of the merger rate
would require the same behavior of the BATSE data for consistency.
As a consequence, the secondary peak at $z_{0.4}=3.8$, although
providing a good fit for peak flux values near the threshold, tends to
depart more and more with the data for higher values of $P$ 
(Fig.3). Therefore, the redshift range related to an assumed initial
burst of star formation in elliptical galaxies is likely to be beyond
the actual limiting redshift for GRBs.
 
Finally, it has to be pointed out that the models
can hardly reproduce the peak flux values of BATSE's faintest bursts (see
also Totani 1997). As shown by Reichard \& M\'esz\'aros (1997), this
feature results from the assumption that GRBs are standard candles.

\begin{figure}[t]
\centerline{
\psfig{file=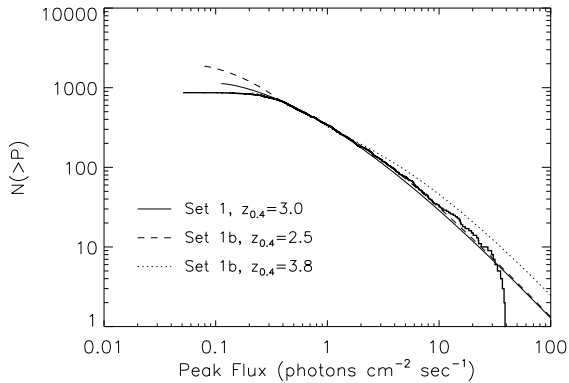,bbllx=95pt,bblly=375pt,bburx=530pt,bbury=695pt,height=5cm}}
\caption[t]{Cumulative counts of bursts $N(>P)$ of the BATSE 3B catalog
(histogram) and model curves
computed for parameters shown in the figure (for 
$t_{gal}(0)$~=~12.16~Gyr) 
}
\end{figure}

\begin{figure}[t]
\centerline{
\psfig{file=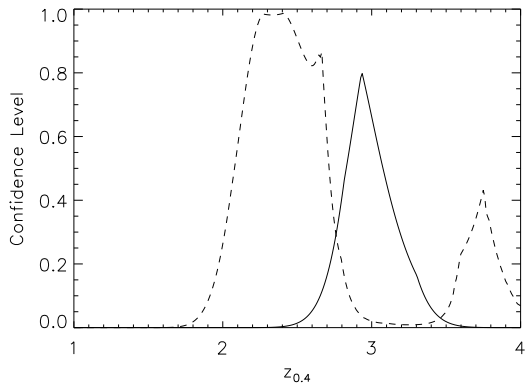,bbllx=95pt,bblly=375pt,bburx=530pt,bbury=695pt,height=5cm}}
\caption[t]{Redshift $z_{0.4}$ necessary to achieve agreement with the
BATSE 3B $\log N - \log P$ distribution.  The solid line corresponds
to set~1 and the dotted line is for set~1b. In both cases, the current
galactic age used is $t_{gal}(0)=12.16$ Gyr.}
\label{fig:KSfig}
\end{figure}

\section{Conclusion}

The computed relative merger rate as a function of redshift is in
agreement with the detected rate of $\gamma$-ray bursts up-to the
limiting redshift of BATSE.  For the models where elliptical galaxies
experience an initial burst of star formation the occurrence rate of
\nsns\ coalescence increases suddenly by more than a factor two at a
redshift of 2.5, which is beyond our current detection limit.  The
synthetic $\log N - \log P$\ distribution is compatible with the
observations down to the completeness limit of $\gamma$-ray
catalogues.  The same is true for models based on observationally
inferred cosmic star formation history (see \eg\ Sahu \etal\ 1997) and
for models based on galactic evolution (Totani 1997). They, however,
predict a different behavior if star bursts occur at high $z$.

The absolute value of the \nsns\ merger rate is found to be
$~\sim$~100 times larger than the GRBs frequency.  Escape from this
conundrum is obtained if the opening angle of the observed phenomenon
is a few degrees, which is consistent with fire-ball models where
leptons are converted into bulk barionic motion (M\'esz\'aros \&
Rees 1992).

If $\gamma$-ray bursts indeed originate from \nsns\ mergers and our
model for star formation is correct, interesting implications follow.
The progenitors of bright GRBs or bursts
at low redshift ($z\la 1.5$) most likely belonged to early-type spiral
galaxies whereas the progenitors of the dimmest bursts and those at
high redshift ($z\ga 2$) were located in elliptical galaxies.  The
majority of the bursts of gravitational waves in this model are
expected to originate from early-type spiral galaxies. If most stars
in the Universe formed in dwarf star-burst galaxies, a substantial
fraction of the parental population may originate from them. Suggested
by current observational data the decline in the star formation rate
beyond $z \approx 1.2$~ (\eg\ Connolly \etal\ 1997), if real, will
show-up as a turn over in the $\log N - \log P$\ distribution for
GRBs.

As noticed by Tutukov \& Yungelson (1994) and confirmed by PZY98
(their Figs.  6 and 8), space velocities of \nsns\ binaries may well
exceed escape velocities of dwarf as well as giant galaxies and they
may travel up to $\sim 1$\,Mpc before coalescence. Thus, a significant
fraction of the sites of GRBs may not be directly associated with star
forming regions.

\acknowledgements We thank Jan van Paradijs for reading the
manuscript.  This work was partially supported  
through NWO Spinoza grant 08-0
to E.~P.~J.~van den Heuvel and RFBR Grant 96-02-16351. L.R.Y
acknowledges hospitality of the Astronomical Institute ``Anton
Pannekoek'' and Meudon Observatory.


\begin{thebibliography}{}
\bibitem[]{}Babul, A., Ferguson, H. C. 1996, ApJ 458, 100
\bibitem[]{}Blinnikov, S.I., Novikov, I.D., Perevodchikova, T.V.,
Polnarev, A.G. 1984, SvAL 10, 177
\bibitem[]{}Bond, H. 1997, IAU Circ. 6654
\bibitem[]{}Bruzual, G.A., Charlot, S. 1993, ApJ 405, 538
\bibitem[]{}Connolly, A.J., Szalay, A.S., Dickinson, M.,
 \etal\ 1997, ApJ 486, L11
\bibitem[]{}Dermer, C.D. 1992, Phys. Rev. Lett. 68, 1799
\bibitem[]{}Dressler, A. 1980, ApJ 236, 351
\bibitem[]{}Efstathiou, G., Ellis, R.S., Peterson B.A. 1988,
MNRAS 232, 431
\bibitem[]{}Gallagher, J. S., Hunter, D., Tutukov, A.V. 1984,
ApJ 284, 544 
\bibitem[]{}Groot, P.J., Galama, T.J., van Paradijs, J., et al. 1997, 
IAU Circ. 6584
\bibitem[]{}Guiderdoni, B., Rocca-Volmerange, B. 1987, A\&A 186, 1
\bibitem[]{}Horack, J.M., Emslie, A.G., Hartmann, D.H. 1995, ApJ 447, 474
\bibitem[]{}Horack, J.M., Mallozzi, R.S., Koshut, T.M. 1996, ApJ 466, 21
\bibitem[]{}Lipunov, V.M.
Nazin S.N., Panchenko I.E., 
\etal\ 1995, A\&A 298, 677
\bibitem[]{}Mao, S., Paczy\'nski, B. 1992, ApJ 388, L45
\bibitem[]{}Meegan, C.A., Pendleton, G.N., Briggs, M.S.
 et al. 1996, ApJS 106, 25
\bibitem[]{}M\'esz\'aros, P., Rees, M.~J. 1992, ApJ 397, 570
\bibitem[]{}Metzger, M.R., Djorgovski, S.G., Kulkarni, S.R.,
 et al.\ 1997, Nat 387, 878
\bibitem[]{}Norris, J.P., Bonnell, J.T., Nemiroff, R.J., \etal\ 1995, ApJ 439, 542
\bibitem[]{}Phinney, E.S. 1991, ApJ 380, L17
\bibitem{}Portegies Zwart, S.F., Verbunt, F. 1996, A\&A 309, 179
\bibitem[]{}Portegies Zwart, S.\ F., Yungelson, L.\ R. 1998, A\&A in
press (astro-ph/9710347)
\bibitem[]{}Reichard, D.E., M\'esz\'aros, P. 1997, ApJ 483 597
\bibitem[]{} Sahu, K.C., Livio, M., Petro, L., \etal\ 1997, ApJ 489, L127
\bibitem[]{}Sandage, A. 1986, A\&A 161, 89  
\bibitem[]{}Tutukov, A.V., Yungelson, L.R. 1994, MNRAS 268, 871
\bibitem[]{}Totani, T. 1997, ApJ 486, L71
\end{thebibliography}
\end{document}